\documentclass{bmcart}
\usepackage{graphicx}
\usepackage{amsmath}
\usepackage{amssymb}

\startlocaldefs

\newcommand{\ket}[1]{\ensuremath{|#1\rangle}}
\newcommand{\bra}[1]{\langle#1|}
\newcommand{\braket}[2]{\langle#1|#2\rangle}

\newcommand{\VL}{V_\textrm{L}}
\newcommand{\VR}{V_\textrm{R}}
\newcommand{\phiL}{\phi_\textrm{L}}
\newcommand{\phiR}{\phi_\textrm{R}}

\newcommand{\fref}[1]{Fig.~\ref{#1}}
\newcommand{\sref}[1]{Sec.~\ref{#1}}
\newcommand{\aref}[1]{App.~\ref{#1}}
\newcommand{\eref}[1]{Eq.~(\ref{#1})}

\newcommand{\Sref}[1]{Section~\ref{#1}}

\endlocaldefs

\graphicspath{{/}{figures/}}

\begin{document}

\begin{frontmatter}

\begin{fmbox}
\dochead{Research}

\title{Spatial non-adiabatic passage using geometric phases}
\author[addressref={oist},corref={oist}]{Albert Benseny}
\author[addressref={ucc}]{Anthony Kiely}
\author[addressref={oist,shanghai}]{Yongping Zhang}
\author[addressref={oist}]{Thomas Busch}
\author[addressref={ucc}]{Andreas Ruschhaupt}

\address[id=oist]{\orgname{Quantum Systems Unit, Okinawa Institute of Science and Technology Graduate University}, \city{904-0495 Okinawa}, \cny{Japan}}
\address[id=ucc]{\orgname{Department of Physics, University College Cork}, \city{Cork}, \cny{Ireland}}
\address[id=shanghai]{\orgname{Department of Physics, Shanghai University}, \city{Shanghai 200444}, \cny{China}}

\end{fmbox}% comment this for two column layout

\begin{abstractbox}

\begin{abstract}
Quantum technologies based on adiabatic techniques can be highly effective, but often at the cost of being very slow. Here we introduce a set of experimentally realistic, non-adiabatic protocols for spatial state preparation, which yield the same fidelity as their adiabatic counterparts, but on fast timescales. In particular, we consider a charged particle in a system of three tunnel-coupled quantum wells, where the presence of a magnetic field can induce a geometric phase during the tunnelling processes. We show that this leads to the appearance of complex tunnelling amplitudes and allows for the implementation of spatial non-adiabatic passage. We demonstrate the ability of such a system to transport a particle between two different wells and to generate a delocalised superposition between the three traps with high fidelity in short times.
\end{abstract}


\end{abstractbox}
\end{frontmatter}

\section{Introduction}

Adiabatic techniques are widely used for the manipulation of quantum states. They typically yield high fidelities and possess a high degree of robustness. One paradigmatic example is stimulated Raman adiabatic passage (STIRAP) in three-level atomic systems~\cite{bergmann_1998}.  STIRAP-like techniques have been successfully applied to a wide range of problems, and in particular, to the control of the centre-of-mass states of atoms in microtraps. This spatial analogue of STIRAP is called spatial adiabatic passage (SAP) and it relies on coupling different spatial eigenstates via a controllable tunnelling interaction~\cite{sap_review}. It has been examined for cold atoms in optical traps~\cite{sap_eckert,sap_atoms,sap_hole,sap_pair,menxi_interf,menxi_am} and for  electrons trapped in quantum dots~\cite{sap_greentree,qd_paspalakisa}. The ability to control the spatial degrees of freedom of trapped particles is an important goal for using these systems in future quantum technologies such as atomtronics~\cite{atomtronics,sap_hole} and quantum information processing~\cite{jaksch_entanglement}. SAP has also been suggested for a variety of tasks such as interferometry~\cite{menxi_interf}, creating angular momentum~\cite{menxi_am}, and velocity filtering~\cite{loiko2014}. It is also applicable to the classical optics of coupled waveguides~\cite{waveguides}.

However, the high fidelity and robustness of adiabatic techniques comes at the expense of requiring long operation times. This is problematic as the system will therefore also have a long time to interact with an environment leading to losses or decoherence. To avoid this problem, we will show how one can speed-up processes that control the centre-of-mass state of quantum particles and introduce a new class of techniques which we refer to as spatial non-adiabatic passage. The underlying foundation for these are shortcuts to adiabaticity (STA) techniques, which have been developed to achieve high fidelities in much shorter total times, for a review see~\cite{sta_review_1,sta_review_2}. Moreover, shortcuts are known to provide the freedom to optimise against undesirable effects such as noise, systematic errors or transitions to unwanted levels~\cite{sta_review_2,sta_refs}.

Implementing the STA techniques for spatial control requires complex tunnelling amplitudes. However, tunnelling frequencies are typically real. To solve this, we show that the application of a magnetic field to a triple well system containing a single charged particle (which could correspond to a quantum dot system~\cite{3qd_review,3qd_ring,3qd_sap,qdots_AB}) can achieve complex tunnelling frequencies through the addition of a geometric phase. This  then allows one to implement a counter-diabatic driving term~\cite{sta_review_1,sta_review_2,rice2003,berry2009,chen2010} or, more generally, to design dynamics using Lewis--Riesenfeld invariants~\cite{LR69}.

The paper is structured as follows. In the next section, we present the model we examine, namely a charged particle in a triple well ring system with a magnetic field in the centre.
In \sref{sec:SAP}, we introduce the spatial adiabatic passage technique in a three-level system and show that making one of the couplings imaginary allows the implementation of transitionless quantum driving. We then show, in \sref{sec:LRI}, how to create inverse-engineering protocols in this system using Lewis--Riesenfeld invariants.
Results for two such protocols, namely transport and generation of a three-trap superposition, are given in \sref{sec:schemes}.
\Sref{sec:1Dsim} presents a more realistic one-dimensional continuum model for the system, where the same schemes are implemented.
Finally, in \sref{sec:conc}, we review and summarise the results.

% ---------------------------------------------------------------
% System model
% ---------------------------------------------------------------

\section{System model}
\label{sec:model}

\begin{figure}
\includegraphics[width=0.4\linewidth]{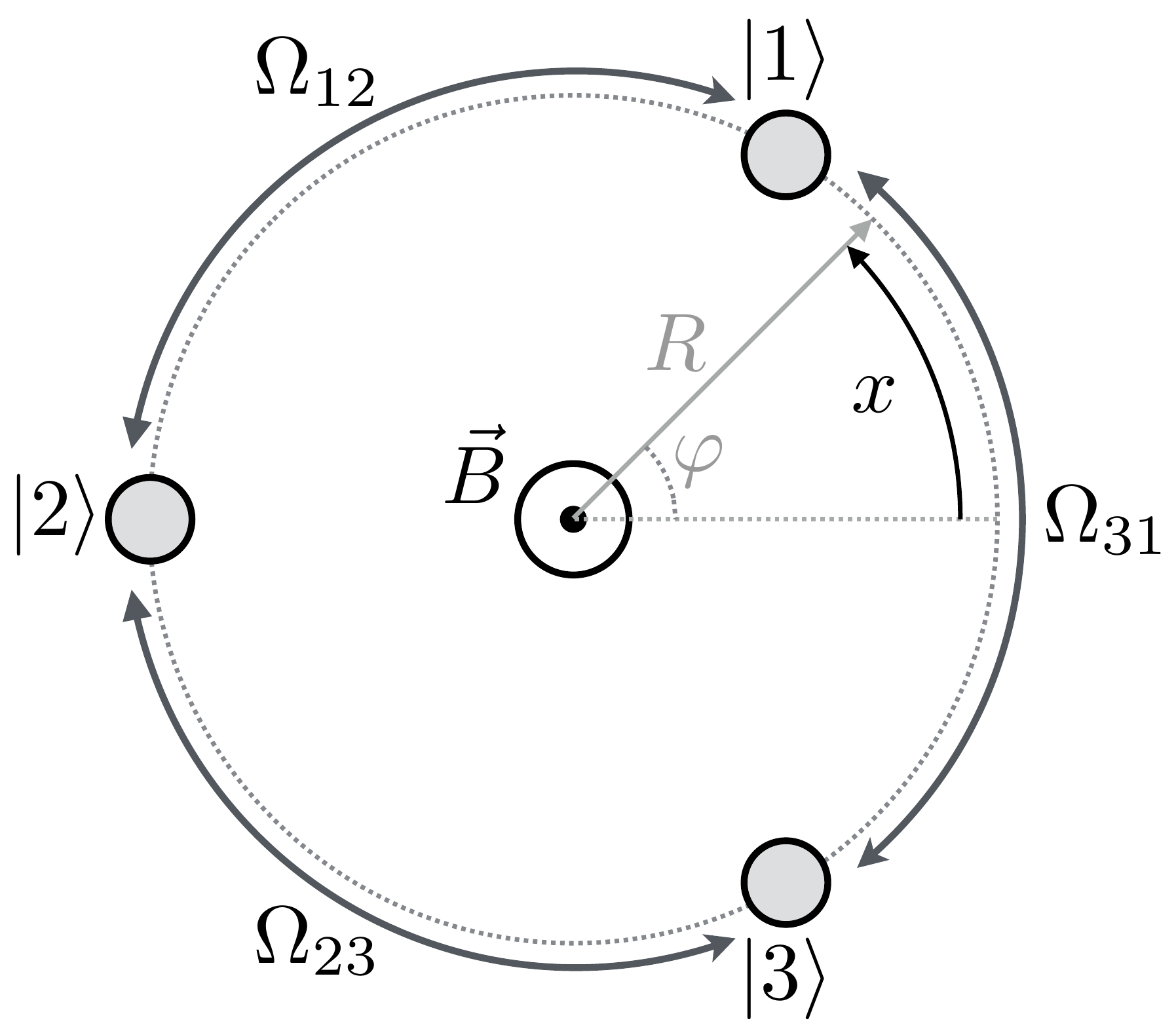}

\caption{\label{fig:sketch}
\csentence{Diagram of the system consisting of three coupled quantum wells  and a localised magnetic field in the centre.} The basis states and the couplings strengths used in the three-level approximation are indicated. The coordinate system for the continuous model in \sref{sec:1Dsim} is also shown.
The distance between two traps along the ring is defined as $l$, so that the total circumference of the ring is $3l$.
}
\end{figure}

We consider a charged particle trapped in a system of three localised potentials, between which the tunnel coupling can be changed in a time-dependent manner.
In order to have coupling between all traps, they are assumed to be arranged along a ring and a magnetic field exists perpendicular to the plane containing the traps, see \fref{fig:sketch}. The particle will initially be located in one of the traps and we will show how to design spatial non-adiabatic passage protocols where a specific final state can be reached within a finite time and with high fidelity.
Such a model could, for example, correspond to an electron trapped in an arrangement of quantum dots, where gate electrodes can be used to change the tunnelling between different traps~\cite{qdot_parameter}.
Another option would be to use ion trapping systems~\cite{wineland_traps}, where ring configurations have been recently demonstrated~\cite{ion_noguchi,ion_ring}.
In these systems, tunnelling of an ion has already been observed (and controlled by manipulating the radial confinement), as well as the 
Aharonov--Bohm phase~\cite{ABeffect} acquired due to the presence of an external magnetic field~\cite{ion_noguchi}.

Let us start by considering the single-particle Schr\"odinger equation
\begin{eqnarray}
\label{eq:TDSE_app}
i\hbar\frac{\partial\psi}{\partial t} = \frac{1}{2m} \left(- i \hbar \nabla - q \vec{A}\right)^2 \psi + V \psi,
\end{eqnarray}
where $m$ and $q$ are the mass and charge of the particle, respectively, and $V$ corresponds to the potential describing the trapping geometry. We assume that the vector potential is originating from an idealised point-like and infinitely long solenoid at the origin (creating a magnetic flux $\Phi_B$) and it is therefore given by $\vec{A} = \frac{\Phi_B}{2 \pi r} \hat{e}_\varphi$ (for $\vec r \neq 0$).  Here $r, \varphi, z$ are cylindrical coordinates and $\hat{e}_\varphi$ is a unit vector in the $\varphi$ direction.

At low energies such a system can be approximated by a three-level (3L) model, where each basis state, $\ket{j}$, corresponds to the localised ground state in one of the trapping potentials
(see \fref{fig:sketch}).
These states are isolated when a high barrier between them exists, but when the barrier is lowered the tunnelling amplitude $\Omega_{jk}$  between states $\ket{j}$ and $\ket{k}$ becomes significant.

The presence of the magnetic field leads to the particle acquiring an Aharonov--Bohm phase~\cite{ABeffect} whenever it moves (tunnels) between two different positions (traps). This phase is given by $\phi_{j,k} = \frac{q}{\hbar}\int_{\vec{r}_j}^{\vec{r}_k}\vec{A}(\vec{r})\cdot d\vec{r}$, where $\vec{r}_j$ is the position of the $j$-th trap, and
for consistency, we always chose the direction of the  path of the integration to be anti-clockwise around the pole of the vector potential (at $\vec r = 0$).
The effects of this phase on the tunnelling amplitudes is given through the Peierls phase factors~\cite{peierls}, 
$\exp\left(i \phi_{j,k} \right)$, and the Hamiltonian for the 3L system can be written as
\begin{equation}
\label{eq:3LevelHamiltonian}
	H = -\frac{\hbar}{2}
		\begin{pmatrix}
			0 & \Omega_{12}e^{i\phi_{1,2}} & \Omega_{31}e^{-i\phi_{3,1}} \\
			\Omega_{12}e^{-i\phi_{1,2}} & 0 & \Omega_{23}e^{i\phi_{2,3}} \\
			\Omega_{31}e^{i\phi_{3,1}} & \Omega_{23}e^{-i\phi_{2,3}} & 0
		\end{pmatrix} .
\end{equation}
Here the $\Omega_{jk}$ are the coupling coefficients in the absence of any vector potential. The total phase around a closed path containing the three traps is then given by
\begin{eqnarray}
\Phi \equiv \phi_{1,2}+\phi_{2,3}+\phi_{3,1} = \frac{q}{\hbar}\oint\vec{A}(\vec{r})\cdot d\vec{l} = \frac{q}{\hbar} \Phi_B,
\end{eqnarray}
and is non-zero due to the pole of the vector potential $\vec A$ at the origin.

To simplify the Hamiltonian \eqref{eq:3LevelHamiltonian} one can use the following unitary transformation, which only employs local phases,
\begin{equation}
	U=\begin{pmatrix}
		1 & 0 & 0\\
		0 & e^{-i\phi_{1,2}} & 0\\
		0 & 0 & e^{-i\left(\phi_{1,2}+\phi_{2,3}\right)} 
		\end{pmatrix} ,
\end{equation}
and transforms the Hamiltonian as
\begin{equation}
H \rightarrow U^{\dagger} H U= - \frac{\hbar}{2}
		\begin{pmatrix}
		0 & \Omega_{12} & \Omega_{31}e^{- i \Phi}\\
		\Omega_{12} & 0 & \Omega_{23}\\
		\Omega_{31}e^{i \Phi} & \Omega_{23} & 0
		\end{pmatrix},
\end{equation}
so that two of the tunnelling amplitudes become real-valued.

A case of particular interest is when $\Phi=\pi/2$, i.e., when the magnetic flux is $\Phi_B = \pi \hbar / 2 q$.
In this case the Hamiltonian  becomes
\begin{equation}
 \label{eq:totalH}
H = -\frac{\hbar}{2}\Big(\Omega_{12} K_1+\Omega_{23} K_2+\Omega_{31} K_3\Big) ,
\end{equation}
where each $K_j$ is a spin 1 angular momentum operator defined as
\begin{equation}
	\label{eq:KSU2}
	K_1=\begin{pmatrix}
	0 & 1 & 0\\
	1 & 0 & 0\\
	0 & 0 & 0
	\end{pmatrix},
\quad 
	K_2=\begin{pmatrix}
	0 & 0 & 0\\
	0 & 0 & 1\\
	0 & 1 & 0
	\end{pmatrix},
\quad 
	K_3=\begin{pmatrix}
	0 & 0 & -i\\
	0 & 0 & 0\\
	i & 0 & 0
	\end{pmatrix},
\end{equation}
satisfying $[K_j, K_k] = i \epsilon_{jkl} K_l$ and $\epsilon_{jkl}$ is the Levi-Civita symbol~\cite{SU2_carroll}. This means that the tunnel coupling  between $|3\rangle$ and $|1\rangle$ becomes purely imaginary.
We will show in the next section that this allows for the implementation of spatial non-adiabatic passage processes by either applying a transitionless quantum driving protocol or by using Lewis--Riesenfeld invariants.

% -----------------------------------------------------------------
% Adiabatic and shortcut transfer in a three-level quantum system
% -----------------------------------------------------------------

\section{Processes in the three-level approximation}
\label{sec:SAP}

\subsection {Adiabatic methods}
A series of spatial adiabatic passage (SAP) techniques have been developed in recent years, which allows one to manipulate and control the external degrees of freedom of quantum particles in localised potentials with high fidelity~\cite{sap_review}.
The standard SAP protocol for the transport of a single particle in a triple well system~\cite{sap_eckert,sap_greentree} is the spatial analogue of the quantum-optical STIRAP technique~\cite{bergmann_1998}.
It involves three linearly arranged, degenerate trapping states, $\ket{j}$ with $j = 1$, $2$ and $3$, that can be coupled through tunnelling by either changing the distance between the traps or lowering the potential barrier between them.
The system in the 3L approximation is described by the Hamiltonian
\begin{equation}
\label{eq:H0}
H_0 = -\frac{\hbar}{2}\left(\Omega_{12} K_1+\Omega_{23} K_2\right),
\end{equation}
which has a zero-energy eigenstate of the form
\begin{equation}
\label{eq:darkstate}
\ket{\lambda_0} = \cos \theta \ket{1} - \sin \theta \ket{3}\quad\text{with}\quad
\tan\theta = \Omega_{12}/\Omega_{23} .
\end{equation}
This state is often called the \textit{dark state} and
SAP consists of adiabatically following $\ket{\lambda_0}$ from $\ket{1}$ (at $t=0$) to $-\ket{3}$ (at a final time $t=T$), effectively transporting the particle between the outer traps one and three. 
This corresponds to changing $\theta$ from $0$ ($\Omega_{23} \gg \Omega_{12}$) to $\pi/2$ ($\Omega_{23} \ll \Omega_{12}$).
Hence in the case of ideal adiabatic following, trap two (located in the middle) is never populated.

\subsection{Transitionless quantum driving \label{Hcd_sect}} 

The main drawback of SAP is that it requires the process to be carried out adiabatically and therefore slowly compared to the energy gap~\cite{sap_review}.
If this requirement is not met, unwanted excitations will lead to imperfect transport.
One way to specifically cancel possible diabatic transitions in STIRAP was discussed in~\cite{Unanyan_OC97} and a general approach for recovering adiabatic dynamics in a non-adiabatic regime is to use shortcuts to adiabaticity, such as transitionless quantum driving~\cite{rice2003,berry2009,chen2010}.
This technique consists of adding a counter-diabatic term to the original Hamiltonian, whose particular form is given as
\begin{equation}
H_{\rm CD} = i \hbar \sum_n \Big( \ket{\partial_t \lambda_n}\bra{\lambda_n} - \braket{\lambda_n}{\partial_t \lambda_n} \ket{\lambda_n}\bra{\lambda_n}\Big),
\end{equation}
where the $\ket{\lambda_n}$ are the eigenstates of $H_0$.
For the reference Hamiltonian in \eref{eq:H0} this gives~\cite{chen2010}
\begin{equation}
	 \label{eq:HCD}
	H_{\rm CD} =- \frac{\hbar \Omega_{31}(t)}{2} K_3,\quad\text{with}\quad 
	\Omega_{31}(t) = 2 \dot{\theta}(t) =2 \left(\frac{\Omega_{23} \dot{\Omega}_{12} - \Omega_{12} \dot{\Omega}_{23}}{\Omega_{12}^2 + \Omega_{23}^2}\right).
\end{equation}
We will see in \sref{sec:trans} how this exact same scheme can also be obtained using Lewis--Riesenfeld invariants.

Shortcuts to adiabaticity have been studied in the context of STIRAP~\cite{chen2010,chen_exp}, i.e., population transfer between internal levels. Its spatial analogue is more challenging as it requires that the additional tunnelling coupling between sites one and three is imaginary (see the definition of $K_3$ in \eref{eq:KSU2}).
However, the system we have presented here is ideal for this, as the system Hamiltonian \eref{eq:totalH} is already equal to the total Hamiltonian $H_0 + H_{\rm CD}$.
Other methods to implement the imaginary coupling could be, for example, the use of artificial magnetic fields~\cite{dalibard_review} or angular momentum states~\cite{polo_am}.

A heuristic but not rigorous explanation of why the coupling needs to be imaginary can be obtained by examining the two ``paths" the particle can take to move from trap one to trap three. The first is
via SAP  and leads to $\ket{1} \to -\ket{3}$ whereas the second is via the direct coupling the shortcut introduces, which leads to $\ket{1} \to i e^{i \Phi} \ket{3}$.
One can then immediately see that for constructive interference of these two terms the phase needs to have the value $\Phi = \pi/2$, which corresponds to the required imaginary coupling between states $\ket{1}$ and $\ket{3}$. It is also interesting to note that the coupling between traps one and three in the shortcut has the form of a $\pi$-pulse 
\begin{equation}
\int_0^T \Omega_{31}(t) dt 
= 2 \int_0^T  \dot{\theta}(t) dt 
= 2 \left[\theta(T) - \theta(0)\right] = \pi .
\end{equation}

\subsection{Invariant-based inverse engineering}
\label{sec:LRI}

Another method of
designing shortcuts to adiabaticity is by means of inverse-engineering using Lewis--Riesenfeld (LR) invariants~\cite{LR69,chen2010b}.
In this section we will briefly review these methods and then apply them to our particular system to both transport the particle and create a superposition between the three wells.

A LR invariant for a Hamiltonian $H(t)$ is a Hermitian operator $I(t)$ satisfying~\cite{LR69}
\begin{equation}
\frac{\partial I}{\partial t}+\frac{i}{\hbar}\left[H,I\right]=0.
\label{eq:LR}
\end{equation}
Since $I(t)$ is a constant of motion it can be shown that it has time-independent eigenvalues. It can be further shown that a particular solution of the Schr\"{o}dinger equation,
\begin{equation}
i\hbar \partial_{t} \ket{\psi(t)} = H(t) \ket{\psi(t)} ,
\end{equation}
can be written as
\begin{equation}
\ket{\psi_{k}(t)}=e^{i \alpha_{k}(t)} \ket{\phi_{k}(t)} ,
\end{equation}
where the $\ket{\phi_{k}(t)}$ are the instantaneous eigenstates of
$H(t)$ and 
\begin{equation}
\alpha_{k}(t)=\frac{1}{\hbar} \int_{0}^t \bra{\phi_{k}(s)} \Big[i\hbar\partial_{s}-H(s)\Big] \ket{\phi_{k}(s)} ds
\end{equation}
are the LR phases. Hence a general solution to the Schr\"{o}dinger
equation can be written as
\begin{equation}
\ket{\psi(t)}=\sum_{k} c_{k} \ket{\psi_{k}(t)},
\end{equation}
where the $c_{k}$ are independent of time.

The idea behind inverse engineering using LR invariants is not to follow
an instantaneous eigenstate of the $H(t)$ as one would in the adiabatic case, but rather follow an eigenstate of $I(t)$ (up to the LR phase).
To guarantee that the eigenstates coincide at the beginning and the end of the process, it is necessary that the invariant and the Hamiltonian commute at these times, i.e.,
\begin{equation}
	\label{eq:commute_bound}
	\left[I(0),H(0)\right]=\left[I(T),H(T)\right]=0.
\end{equation}
One is then free to choose how the state evolves in the intermediate time and
once this is fixed, \eref{eq:LR} determines how the Hamiltonian should vary with time to achieve those dynamics.

A LR invariant for a three-level system described by \eref{eq:totalH} can be written as
\begin{equation}
I = -\sin\beta\sin\alpha K_1-\sin\beta\cos\alpha K_2+\cos\beta K_3 .
\end{equation}
where $\alpha$ and $\beta$ are time dependent functions which must fulfil the following relations (imposed by \eref{eq:LR})
\begin{align}
\label{inveq1}
\dot{\alpha} &= \frac{\Omega_{12} \sin\alpha  + \Omega_{23} \cos\alpha }{2 \tan\beta} + \frac{\Omega_{31}}{2}, \\
\label{inveq2}
\dot{\beta} &= \frac{1}{2} (\Omega_{23} \sin\alpha - \Omega_{12} \cos\alpha).
\end{align}
The eigenstates of this invariant are 
\begin{align}
\ket{\phi_{0}(t)} &= \left(\begin{array}{c}
-\sin\beta\cos\alpha\\
-i\cos\beta\\
\sin\beta\sin\alpha
\end{array}\right), \\
\ket{\phi_{\pm}(t)} &= \frac{1}{\sqrt{2}}\left(\begin{array}{c}
\cos\beta\cos\alpha\pm i\sin\alpha\\
-i\sin\beta\\
-\cos\beta\sin\alpha\pm i\cos\alpha
\end{array}\right),
\end{align}
with respective eigenvalues $\mu_{0}=0$ and $\mu_{\pm}=\pm 1$.
One solution of the time-dependent Schr\"odinger equation is then given by $\ket{\Psi(t)} = \ket{\phi_0 (t)}$ as the corresponding LR phase is zero in this case.
Note that this invariant is a generalisation of the invariant considered in Ref.~\cite{Chen_PRA2012} where a third coupling $\Omega_{31}$ was not taken into account.

After fixing the boundary conditions using \eref{eq:commute_bound}, one is free to choose the functions $\alpha(t)$ and $\beta(t)$.
Moreover, in this case, one is also free to directly choose the function $\Omega_{31}$.
By inverting Eqs.~\eqref{inveq1} and \eqref{inveq2}, the other coupling coefficients 
are then given by
\begin{align}
\label{rabi_12}
\Omega_{12} &= 2 \dot\alpha \sin\alpha \tan\beta - 2 \dot\beta \cos\alpha - \Omega_{31} \sin\alpha \tan\beta , \\
\label{rabi_23}
\Omega_{23} &= 2 \dot\alpha \cos\alpha \tan\beta + 2 \dot\beta \sin\alpha - \Omega_{31} \cos\alpha \tan\beta .
\end{align}

% -----------------------------------------------------------------
% Manipulation schemes in the three-level approximation
% -----------------------------------------------------------------

\section{Examples of spatial non-adiabatic passage schemes}
\label{sec:schemes}

In the following we will discuss two examples of spatial non-adiabatic passage derived from LR invariant based inverse engineering in the 3L approximation.
The first one is the transport between two different traps, which is shown to be equivalent to the transitionless quantum driving method from \sref{sec:SAP} in some cases. 
The second scheme will create an equal superposition of the particle in all three traps.

\subsection{Transport}
\label{sec:trans}

The first example of control we examine is the population transfer determined by
\begin{equation}
\ket{\Psi(0)}=\ket{1} \quad\rightarrow\quad
\ket{\Psi_{\rm target}} = \Psi(T) =- \ket{3},
\end{equation}
which was considered in the optical regime in Ref.~\cite{chen2010}.
This can be achieved by choosing auxiliary functions that fulfil the boundary conditions
\begin{equation}
\beta(0)= \beta(T)= - \frac{\pi}{2}, \quad
\alpha(0)=0, \quad\text{and}\quad
\alpha(T)=\frac{\pi}{2}.
\end{equation}
The experimentally required tunnelling frequencies are then explicitly given by Eqs.~\eqref{rabi_12} and \eqref{rabi_23}.

For the special choice of $\beta(t) = -\pi/2$, one can show that $\braket{2}{\Psi(t)} = 0$ for all times, i.e.~trap two is never occupied during the process. 
This choice then results in
\begin{equation}
\tan\alpha = \frac{\Omega_{12}}{\Omega_{23}}
\quad\text{and}\quad 
\label{eq:omega31}
\Omega_{31} = 2\dot{\alpha}.
\end{equation}
By identifying $\alpha$ with $\theta$ (see \eref{eq:darkstate}) one can immediately see that this is the same pulse as in the STA scheme derived in \sref{Hcd_sect}.

\begin{figure}
\includegraphics[width=0.7\linewidth]{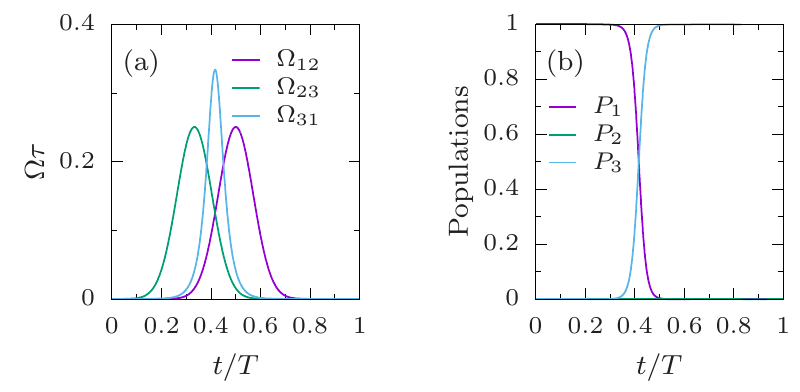}

\caption{\label{fig:sta_3LM}
\csentence{Spatial non-adiabatic passage transport in the 3L approximation.} $T/\tau=100$ for $\Omega_0 \tau = 0.25$.
(a) Modulus of the tunnelling amplitudes.
(b) Evolution of the populations $P_{i}=\left|\braket{i}{\Psi(t)}\right|^{2}$.
The time unit $\tau$ is defined as $\tau = m l^2/\hbar$.}
\end{figure}

The transport scheme can be implemented by the choosing the counterintuitive SAP pulses $\Omega_{12}$ and $\Omega_{23}$ to have a Gaussian profile~\cite{sap_review} 
\begin{align}
\Omega_{12}(t) &= \Omega_0 \exp\left[-100 \left(t/T - 1/2\right)^2\right], \\
\Omega_{23}(t) &= \Omega_0 \exp\left[-100 \left(t/T - 1/3\right)^2\right],
\end{align}
and then calculating $\Omega_{31}$ from Eq.~\eqref{eq:omega31}.
The resulting pulses and associated dynamical populations are shown in \fref{fig:sta_3LM}.
As expected the system follows exactly the dark state, transferring the population between states $\ket{1}$ and $\ket{3}$ without populating state $\ket{2}$.

\begin{figure}
\includegraphics[angle=0,width=0.7\linewidth]{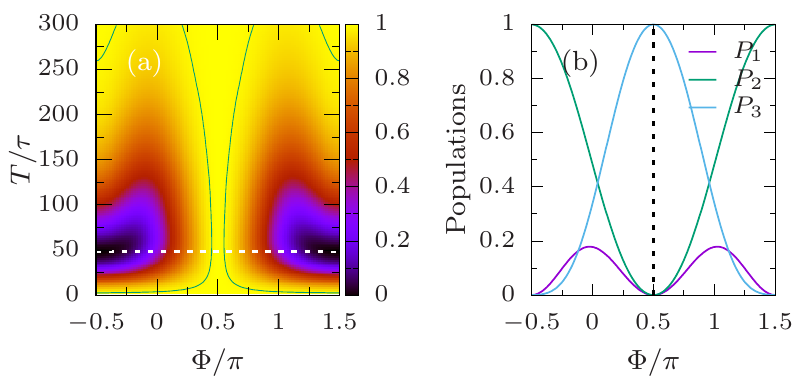}

\caption{\label{fig:STAphase} \csentence{Transport process $\ket{1} \to -\ket{3}$ in the 3L approximation.}
(a) Fidelity as a function of the total time and the total magnetic phase traversing the system. The green contour line is defined by $P_{3}=99\%$.
(b) Probabilities of population in each of the traps for $T/\tau=48$ (indicated by a dashed white line in (a)) as a function of the total magnetic phase traversing the system. The dashed black line indicates the optimal value of the phase $\Phi=\pi/2$}
\end{figure}

The fidelity of the transport process as a function of the total time and the phase $\Phi$ generated by the magnetic field is shown in \fref{fig:STAphase}(a).
Transport can be seen to occur with perfect fidelity for any value of the total time if the phase takes the appropriate value $\Phi = \pi/2$.
It can also be seen that  the shortcut is successful for any value of the phase in the limit of very short or very long times. The  latter one is not surprising, as $\Omega_{31}$
 can be neglected in the adiabatic limit, and hence its phase becomes irrelevant.
A similar effect occurs for short total times, where the roles are reversed.  In this limit $\Omega_{31}$ is the largest of all three couplings, and hence the phase relation between it and the other couplings becomes inconsequential.
As $\Omega_{31}$ is a $\pi$ pulse, perfect population transfer in this regime can be achieved regardless of the phase.

However, in order to maintain this pulse area, a strong coupling is required for very short processes, as the strength of $\Omega_{31}$ is inversely proportional to $T$. 
This sets a bound on how fast this scheme can be implemented, as any physical implementation will have a maximum tunnelling amplitude.
Setting the maximum value of $\Omega_{31}$ to $0.25/\tau$, the minimum process times $T$ to achieve fidelities above $99\%$ are approximately $880 \tau$ for SAP and $100 \tau$ for the shortcut scheme.
These times are similar to the ones achievable in a spin-dependent transport scheme recently presented by Masuda \etal~\cite{masuda}, however the setup in their work requires four traps and a constant and an AC magnetic field.

It is worth noting that this system also allows for the possibility of measuring the magnetic flux $\Phi_{B}$, as the amount of transferred population oscillates as a function of the total phase $\Phi$, which is directly related to the magnetic flux as $\Phi=\frac{q}{\hbar}\Phi_{B}$.
 As an example we show  the occupation probabilities for $T/\tau = 48$ in each trap at the end of the process as a function of the phase  in \fref{fig:STAphase}(b) .
One can see that the populations strongly depend on the phase and over a large range of values one can therefore determine the magnetic flux.
The exact relationship between the probabilities and the magnetic flux differs for different total times $T$.

\subsection{Creation of a three-trap superposition}
\label{sec:super}

\begin{figure}
\includegraphics[width=0.7\linewidth]{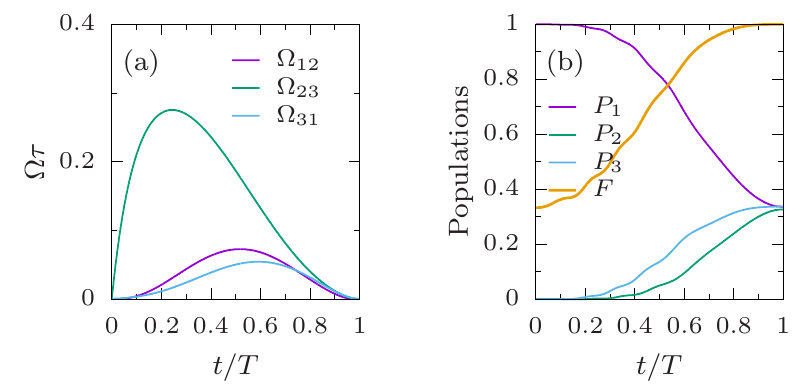}

\caption{\label{fig:sup_3LM}
\csentence{Spatial non-adiabatic superposition scheme $ \ket{1}  \to \frac{1}{\sqrt{3}} (\ket{1} - i \ket{2} - \ket{3})$ in the 3L approximation}. $T/\tau=400$.
Sub-figures are the same as in \fref{fig:sta_3LM} and the fidelity shown in (b) is defined as $F=\left|\braket{\Psi_{\rm target}}{\Psi(t)}\right|^{2}$.}
\end{figure}

The second scheme we discuss highlights the generality of the LR invariant based method. In this scheme we create an equal superposition state between the particles being in all three traps, which means that the initial and target states are 
\begin{equation}
\label{eq:sup_state}
\ket{\Psi(0)}  = \ket{1}\quad\rightarrow\quad
\ket{\Psi_{\rm target}} = \ket{\Psi(T)} = \frac{1}{\sqrt{3}} (\ket{1} - i \ket{2} - \ket{3}).
\end{equation}
This can be realised by imposing the boundary conditions
\begin{align}
&\beta(0)=-\frac{\pi}{2}, \quad
&&\beta(T)=-\arctan \sqrt{2}, \\
&\alpha(0)=0, \quad
&&\alpha(T)=\frac{\pi}{4},
\end{align}
on the auxiliary functions.
A simple ansatz which fulfils these boundary conditions is a fourth order polynomial for $\beta(t)$ and third order polynomials for $\alpha(t)$ and $\Omega_{31}(t)$.
The pulses are then obtained from Eqs. \eqref{rabi_12} and \eqref{rabi_23} and their form is shown in \fref{fig:sup_3LM}(a). From \fref{fig:sup_3LM}(b) it can be seen that this choice creates the target state at the final time with perfect fidelity.

% -----------------------------------------------------------------
% Continuum model
% -----------------------------------------------------------------

\section{Spatial non-adiabatic passage in the continuum model}
\label{sec:1Dsim}

While the 3L approximation discussed above gives a clear picture of the physics of the system, it does not include effects such as excitations to higher energy states that can occur during the process.
We will therefore in the following test the approximation by numerically integrating the full Schr\"odinger equation in real space.
For this, we will consider traps that are narrow enough to limit the system dynamics to an effectively one-dimensional setting along the azimuthal coordinate, $x = \varphi R$, i.e., around a circle of radius $R$, see \fref{fig:sketch}.
Moreover, we will assume that the magnetic field is characterised by a vector potential in the azimuthal direction, $\vec{A} = A \hat{e}_\varphi$.

We are therefore dealing with a one-dimensional system of length $2 \pi R$ with periodic boundary conditions, whose dynamics are described by the following Schr\"odinger equation 
\begin{equation}
\label{eq:TDSE}
i \hbar \frac{\partial \psi}{\partial t} = \frac{1}{2m} \left(- i \hbar \frac{\partial}{\partial x} - q A\right)^2 \psi + V(x) \psi .
\end{equation}
We assume a constant vector potential throughout the dynamical part of the protocols, as any time-varying vector potential would produce an unwanted force due to the electric field $\vec{E} = - \partial_t \vec{A}$.

\begin{figure}
\includegraphics[width=0.7\linewidth]{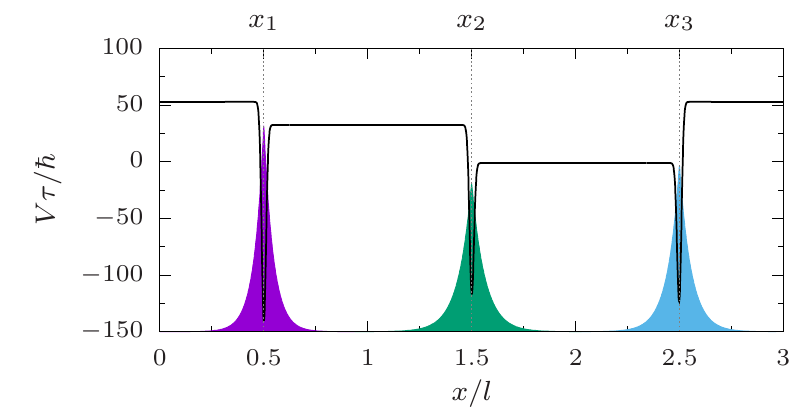}

	\caption{\label{fig:potential}
\csentence{Schematic of the potential used in the numerical simulations (black line) with the localised states in each trap (coloured areas).}
The Gaussian shape of the traps is exaggerated here for clarity.
}
\end{figure}

In order to be able to apply a well-defined phase we model the trapping sites as highly localised point-like potentials of depth $\epsilon_j$ at the positions $x_j = j l - l/2$ (see \fref{fig:potential}).
They are separated by square barriers of heights $V_{jk}(t)$ (and length $l$), giving a total potential
\begin{equation}
	\label{eq.Vtraps}
	V(x,t) =
	- \sum_{j=1}^3 \epsilon_j(t) \delta(x - x_j)+
	\begin{cases}
	V_{31}(t) & \textrm{if} \quad 0 < x < x_1, \\
	V_{12}(t) & \textrm{if} \quad x_1 < x < x_2, \\ 
	V_{23}(t) & \textrm{if} \quad x_2 < x < x_3, \\ 
	V_{31}(t) & \textrm{if} \quad x_3 < x < 3l .
	\end{cases}
\end{equation}
Since point-like potentials are difficult to implement numerically, in the simulations below they are implemented as narrow Gaussians. 
It is important to note that this model is not designed to give realistic estimates for the fidelities or exactly reproduce the dynamics of the 3L approximation.
It is a toy model to validate the basic underlying processes and show that our schemes also make sense in the continuum.

\begin{figure}
\includegraphics[width=0.7\linewidth]{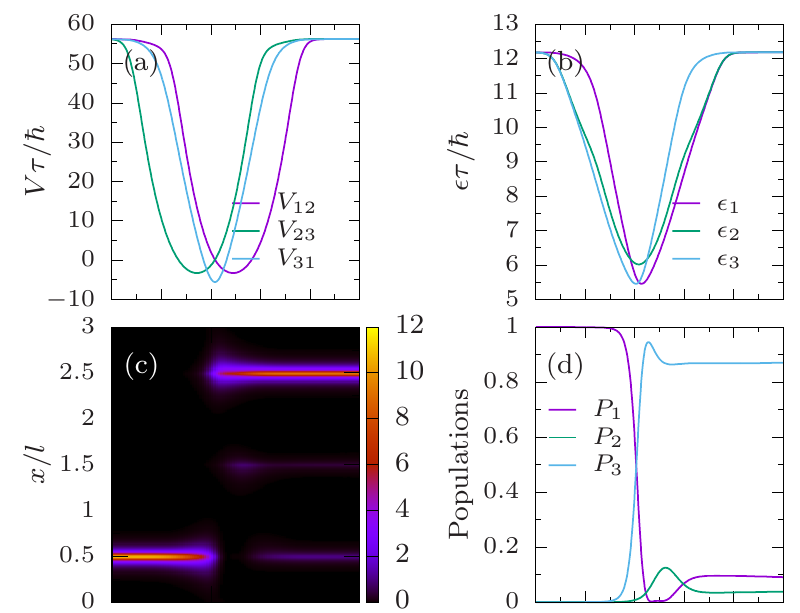}
\includegraphics[width=0.7\linewidth]{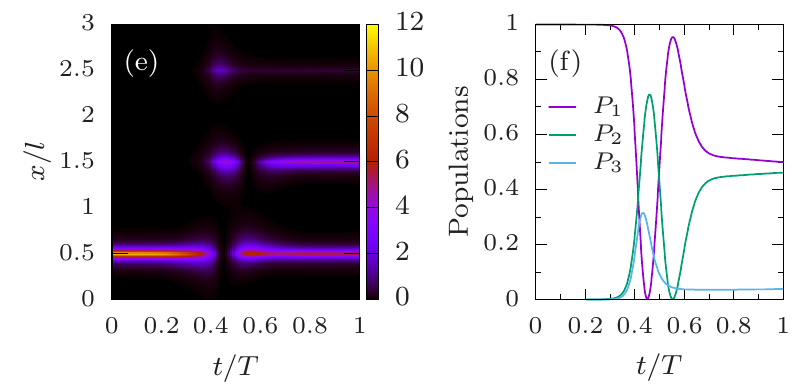}

\caption{\label{fig:sta_1DM}
\csentence{Spatial non-adiabatic transport process in the continuum model.} $T/\tau=100$.
(a,b) Barrier heights and trap depths obtained by mapping the couplings in \fref{fig:sta_3LM}(a).
(c) Evolution of the particle density $|\psi(x,t)|^2$.
(d) Corresponding populations $P_{i}=\left|\braket{i}{\Psi(t)}\right|^{2}$ in each trap and of the target state.
(e,f) are the same as (c,d) but with the magnetic flux flowing in the opposite direction.
The width of the Gaussian traps is $10^{-4}l$.}
\end{figure}

\begin{figure}
\includegraphics[width=0.7\linewidth]{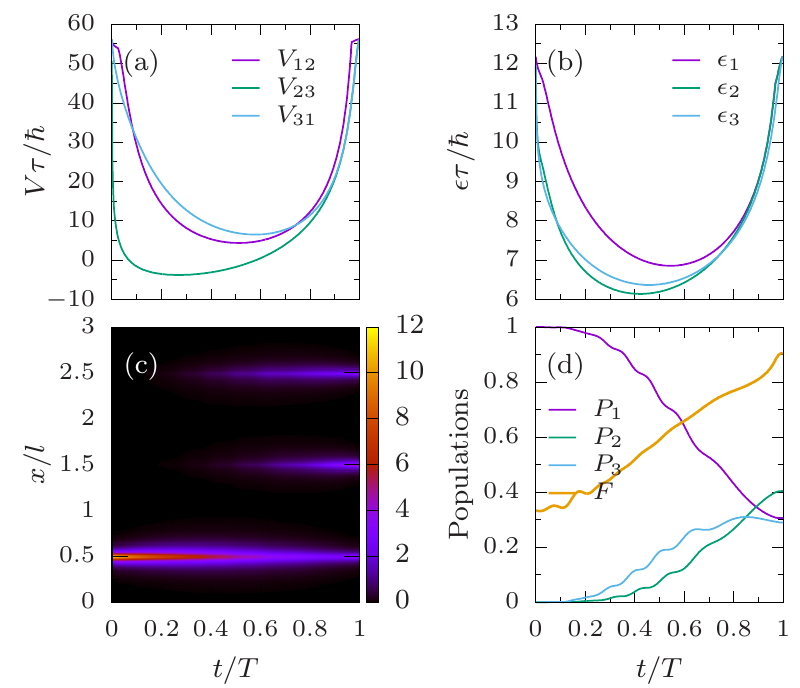}

\caption{
\label{fig:sup_1DM}
\csentence{Same as \fref{fig:sta_1DM}(a-d) but for the spatial non-adiabatic superposition scheme given in \eref{eq:sup_state} in the continuum model.} $T/\tau=400$.
$F=\left|\braket{\psi_{\rm target}}{\psi(t)}\right|^{2}$ is the fidelity of the process.
}
\end{figure}

As mentioned above, the tunnelling amplitudes $\Omega_{jk}(t)$ in the 3L approximation are related to the barrier heights $V_{jk}(t)$ of the continuum model, see \aref{app:mapp}.
However, changing the barrier heights in order to achieve tunnelling will also affect the energies of the localised states in the neighbouring traps.
Therefore, in order to reproduce the resonance of the 3L approximation (where the diagonal elements of the Hamiltonian are always zero) in the continuum model,
the depths of the delta potentials $\epsilon_j$ have to be adjusted as the barriers heights change, see \fref{fig:potential}.
Finally, to map the barrier heights $V_{jk}$ and trap depths $\epsilon_j$ parameters of the continuum model to the tunnelling amplitudes $\Omega_{jk}$ of the 3L approximation, we
numerically calculate the overlaps of neighbouring delta-trap eigenstates.

Results for transport of a particle using the shortcut scheme described in \sref{sec:trans} are shown in \fref{fig:sta_1DM}
 and the barrier heights and trap depths used to match the pulses given in \fref{fig:sta_3LM}  are shown in Figs.~\ref{fig:sta_1DM}(a,b). 
The probability density during the process can be seen in \fref{fig:sta_1DM}(c) and the populations in each trap are given in  \fref{fig:sta_1DM}(d). While the process is not perfect, one can see that the particle is transported to the final trap with a fidelity of 87\%.
The effect of the magnetic field can be seen in Figs.~\ref{fig:sta_1DM}(e,f), where we show results for the same process but with an inverted magnetic field (using a total phase of $\Phi = -\pi/2$).
In this case the interference between the adiabatic and shortcut paths is destructive, and almost no population ends up in the final trap.

The results for the creation of the superposition state discussed in \sref{sec:super} are shown in \fref{fig:sup_1DM}. The observed dynamics are very similar to the one in the 3L approximation and the process reaches a final fidelity of the target state of 91\%.

Since the continuum model has many more degrees of freedom than the 3L model, it is not surprising that the fidelities obtained are lower. Nevertheless, the basic functioning of our spatial non-adiabatic techniques is clearly established from the calculations shown above. Optimising the fidelity in the continuum is an interesting task which, however, goes beyond the scope of the current work.

% -----------------------------------------------------------------
% Conclusions
% -----------------------------------------------------------------

\section{Conclusions and outlook}
\label{sec:conc}

We have shown how complex tunnel frequencies in single-particle systems allow one to develop spatial non-adiabatic passage techniques that can lead to fast and robust processes for quantum technologies.
In particular, we have discussed the case of a single, charged particle in a microtrap environment.
The complex tunnelling couplings are obtained from the addition of a constant magnetic field,
and have allowed us to generalise adiabatic state preparation protocols beyond the usual spatial adiabatic passage techniques~\cite{sap_review}.
This demonstrates that non-adiabatic techniques can be as efficient as their adiabatic counterparts, without requiring the long operation times. 

In particular, we have discussed the implementation of the counter-diabatic term for spatial adiabatic passage transport via a direct coupling of all the traps.
This was, in a second step, generalised to a flexible and robust method for preparing any state of the single-particle system by using Lewis--Riesenfeld invariants.
As an example, we have shown that an equal spatial superposition state between the three wells can be created on a short time scale.
Finally, we have presented numerical evidence that spatial non-adiabatic processes work also in a one-dimensional toy model by introducing a mapping between the discrete three-level approximation and a continuum model.

While in this work we have focused on a three-trap system, an interesting extension would be to investigate similar schemes in larger systems, or in different physical settings (for example, superconducting qubits~\cite{chiral}).
Often, if the transitionless quantum driving technique is directly applied to complex quantum systems,
the additional counter-adiabatic terms become very complicated, hard to implement or even unphysical.
Nevertheless, the steps outlined in our work (using a few-level approximation, applying the shortcut technique, and then mapping everything back to a continuous model) can in principle be applied to any trap configuration.
These steps might lead to schemes which are much easier to implement experimentally than the direct application of the transitionless quantum driving.
However, each of these generalised configuration would need to be studied on an individual basis.

It would also be very interesting to see the effect of interactions in this system.
For very strong interactions such that double occupancy of a site is suppressed and a single empty site is present,
one might expect to observe similar dynamics but for the empty site~\cite{sap_hole}.
In this case, spatial non-adiabatic ideas can be straightforwardly transferred.
For intermediate interaction strengths (but stronger than the tunnelling couplings),
repulsively-bound pair processes have been shown to dominate the dynamics and single-particle-like dynamics can be recovered for the pair~\cite{sap_pair,qd_doublon}.
In this case the presented techniques might be extended for a particle pair.

Finally, it is also worth noting that these complex tunnelling couplings we introduce can be used to implement techniques based on composite pulses~\cite{composite_pulses}.

\begin{appendix}

\section{Mapping between three-level approximation and continuum model}
\label{app:mapp}

In this appendix we give more details on how to connect the parameters of the three-level (3L) approximation and the continuum model. For clarity,  we set $\hbar=m=1$ in the following.
Let us first recall the eigenfunctions of a single asymmetric delta potential given by
\begin{equation}
V(x)=-\epsilon \delta(x)+\left\{ \begin{array}{cc} \VL & x<0 \\ \VR & x \geq 0. \end{array}
\right . 
\end{equation}
This potential has only one bound state
(as long as $2 \epsilon^2 > |\VL - \VR|$)
which is of the form
\begin{equation}
\psi(x)=\left\{ \begin{array}{cc} \phiL(x) & x<0 \\ \phiR(x) & x \geq 0 \end{array}
\right . ,
\end{equation}
where
\begin{align}
\phiL(x) &= \exp \left[\frac{\left(2 \epsilon^{2}+\VL-\VR\right)x}{2 \epsilon}\right] \frac{\sqrt{4 \epsilon^{4}-\left(\VL-\VR\right)^{2}}}{2 \epsilon^{3/2}}, \\
\phiR(x) &= \exp \left[-\frac{\left(2 \epsilon^{2}+\VR-\VL\right)x}{2 \epsilon}\right] \frac{\sqrt{4 \epsilon^{4}-\left(\VL-\VR\right)^{2}}}{2 \epsilon^{3/2}},
\end{align}
with an energy
\begin{equation}
E=-\frac{4 \epsilon^{4}+\left(\VR-\VL\right)^{2}-4 \epsilon^{2}\left(\VL+\VR\right)}{8 \epsilon^{2}}.
\end{equation}

In our work we use these  eigenstate as the localised basis states in each of the three delta trap potentials.
For example, the basis state $\psi_1(x)$ for the first trap can be constructed from the substitutions $\epsilon \rightarrow \epsilon_1$, $\VL \rightarrow V_{13}$, $\VR \rightarrow V_{12}$ and $x \rightarrow x-l/2$ (see \fref{fig:potential}) and the states $\psi_2(x)$ and $\psi_3(x)$ for the other two wells can be obtained in a similar manner.
While choosing a basis for the system this way does not necessarily lead to an orthogonal basis set, we have checked numerically that the states are approximately orthogonal at all times during our simulations.
This allows us to approximate the Hamiltonian associated with \eref{eq:TDSE} as \eref{eq:totalH}.

The couplings constants between each pair of neighbouring two traps can be determined by calculating the overlap between the two respective trap states in the barrier region between them, 
i.e.,
\begin{align}
\Omega_{12}&\approx -2 \int_{l/2}^{3l/2} \psi_1(x) \left[-\frac{1}{2}\partial_{x}^{2} \psi_2(x)+V_{12} \psi_2(x)\right] dx, \\
\Omega_{23}&\approx -2 \int_{3l/2}^{5l/2} \psi_2(x) \left[-\frac{1}{2}\partial_{x}^{2} \psi_3(x)+V_{23} \psi_3(x)\right] dx, \\
\Omega_{31}&\approx -2 \int_{5l/2}^{7l/2} \psi_1(x-3l) \left[-\frac{1}{2}\partial_{x}^{2} \psi_3(x)+V_{31} \psi_3(x)\right] dx.
\end{align}

Similarly, the on-site energies (or diagonal elements) are approximated by considering only the regions for which the basis states are significant, e.g.,
\begin{align}
E_2 \approx & \int_{l/2}^{3l/2} \psi_2(x) \left[-\frac{1}{2}\partial_{x}^{2} \psi_2(x)+V_{12} \psi_2(x)\right] dx \nonumber \\ &+ 
\int_{3l/2}^{5l/2} \psi_2(x) \left[-\frac{1}{2}\partial_{x}^{2} \psi_2(x)+V_{23} \psi_2(x)\right] dx
\end{align}
and correspondingly for the energies $E_1$ and $E_3$. We can then tabulate sets of values for the barrier heights, $\{ V_{12},V_{23},V_{13}\}$, trap depths, $\{ \epsilon_1,\epsilon_2,\epsilon_3\}$, and tunnelling amplitudes, $\{ \Omega_{12},\Omega_{23},\Omega_{13}\}$,
such that the energies all match a fixed reference value, i.e., $E_1=E_2=E_3=E_{0}$ where $E_{0}$ is fixed to some constant value.
Since for a given protocol the required tunnelling amplitudes are known, we can finally numerically invert the table in order to determine how the barrier heights and trap depths have to vary in time.

\end{appendix}

\begin{backmatter}

\section*{Acknowledgements}
This work has received financial support from Science Foundation Ireland under the International Strategic Cooperation Award Grant No. SFI/13/ISCA/2845 and the Okinawa Institute of Science and Technology Graduate University.
We are grateful to David Rea for useful discussion and commenting on the manuscript.

% ------------- References -----------------

\end{backmatter}

\end{document}